\documentclass[a4paper,12pt]{article}
\usepackage[utf8]{inputenc}
\usepackage[T1]{fontenc}       
\usepackage{a4wide}
\usepackage{graphicx}
\usepackage[numbers,super,square]{natbib}
\usepackage{amssymb,amsmath}

\begin{document}

\noindent
\textbf{Article type: Full Paper}

\noindent
{\large
  \textbf{A deep learning approach to identify local structures in atomic-resolution transmission electron microscopy images}
  \par}

\noindent
\textit{Jacob Madsen, Pei Liu, Jens Kling, Jakob Birkedal Wagner,
  Thomas Willum Hansen, Ole Winther, Jakob Schiøtz$^\ast$}

\vspace{1cm}

\noindent
Jacob Madsen, Jakob Schiøtz\\
Center for Atomic-scale Materials Design (CAMD)\\
Department of Physics, Technical University of Denmark\\
2800 Kgs. Lyngby, Denmark\\
E-mail: schiotz@fysik.dtu.dk\\
Pei Liu, Jens Kling, Jakob Birkedal Wagner, Thomas Willum Hansen\\
Center for Electron Nanoscopy (CEN)\\
Technical  University of Denmark\\
2800 Kgs. Lyngby, Denmark\\
Ole Winther\\
Department of Applied Mathematics and Computer
Science\\
Technical University of Denmark\\
2800 Kgs. Lyngby, Denmark


\begin{abstract}
Recording atomic-resolution transmission electron microscopy (TEM)
images is becoming increasingly routine. A new bottleneck is then
analyzing this information, which often involves time-consuming
manual structural identification. We have developed a deep
learning-based algorithm for recognition of the local structure in
TEM images, which is stable to microscope parameters and noise. The
neural network is trained entirely from simulation but is capable
of making reliable predictions on experimental images.  We apply the
method to single sheets of defected graphene, and to metallic
nanoparticles on an oxide support.
\end{abstract}

\section{Introduction}

With the developments in transmission electron microscopes that has
occurred over the last decade, it has become increasingly common to
record and store large amounts of TEM data, often in the form of image sequences.  
This development has been accelerated by the advent of
faster and more sensitive detectors such as the direct detection
camera \cite{McMullan:2014jn}; but also by the development of the Environmental
TEM, where it becomes possible to study how e.g. nanoparticles respond
to reaction gases in real time.\cite{Wagner:2012ec}

As the amount of TEM data available increases, it becomes important to have efficient and automated analysis
tools. In many applications, accurate identification and classification
of local structure is a crucial first step in deriving useful
information from atomic-resolution images. Examples include
characterizing the distribution of dopants \cite{Meyer:2011ez} and
defects \cite{Meyer:2008hj}, \emph{in situ} imaging of phase transformations
\cite{He:2014ew}, structural reordering during materials growth
\cite{Nagao:2015jm,Li:2017ke}, dynamic surface phenomena
\cite{Schneider:2014do} and identification of chemical phases in
nanoparticles \cite{Hussaini:2017hr}.

Analysis methods such as Geometric Phase Analysis (GPA)
\cite{Hytch:1998gr} are based on the local symmetry and periodicity, and has
been very successful at extracting structural information in many
regular structures, including identifying defects, strain and phase
boundaries.\cite{Zhu:2013er}.  However, GPA typically has difficulties 
analyzing e.g.\ surfaces, where the periodicity changes
rapidly.\cite{Madsen:2017ip}

Real space approaches typically either rely on direct identification
of atomic positions by finding local intensity extrema,\cite{Bierwolf:1993ic, Galindo:2007jj} or on
direct comparison with a template\cite{Zuo:2014fw}.  However, these
methods are in general not able to compete with a trained human
expert.  The difficulties arise in part due to the phase contrast
nature of high resolution TEM, which makes the image extremely sensitive to small
changes in the defocus, necessitating human intervention in the image
analysis. When analyzing image sequences, it may even be
necessary to adjust the image analysis tools to each frame, as small
rotations, vibrations and thermal drift can modify the appearance from
one frame to another.  These difficulties are compounded by the low
signal-to-noise ratio resulting from using the smallest possible
electron dose to minimize beam damage to the sample.

Recently, convolutional neural networks and related deep-learning
methods have demonstrated excellent performance in visual recognition
tasks, including particle detection\cite{Zhu:2016qoi} and automatic
segmentation of brain images from cryo-electron microscopy
images.\cite{Quan2016} Kirschner and Hillebrand have published a method for
predicting defocus and sample thickness\cite{Kirschner:2000kb}, and
Meyer and Heindl have used
neural networks to reconstruct the exit wave function from off-axis
electron holograms\cite{Meyer:2008je}.

Whereas deep learning methods have recently been proposed to analyse 
Scanning TEM (STEM) images\cite{Ziatdinov:2017ct}, but have to our knowledge not yet been
used to analyze the atomic structure in high resolution HRTEM images.  In this
article, we describe a CNN based method for classifying atomic
structures in TEM, and demonstrate that it can be applied to single
layers of graphene, as well as to supported metallic nanoparticles.
Under good circumstances, the method can be generalized to identify
chemical species and to identify the height of atomic columns.

\section{Methods}
\label{sec:methods}

The task of identifying atoms in atomically resolved TEM images is a
special case of a general problem in image analysis.  The task is to
identify instances of a set of structures, and assigning class labels
$\{c_n\}$ and Cartesian coordinates $\{(x_n, y_n)\}$ to each of them.
In the simplest case, there is only one class (``atom''), but the
analysis can be extended to identify specific structures of atoms;
atom columns of various sizes; vacancies; etc.  The neural network will
be looking for a predefined set of $N_c$ labels, $C = \{c_0, c_1,
\ldots, c_{N_c}\}$ and will initially assign a probability for each
possible label.  The choice of how to categorize the structures is
problem specific, and typically depends on how the researcher derives
meaning from the image.

\begin{figure*}[t]
  \centering
  \includegraphics[width=16cm]{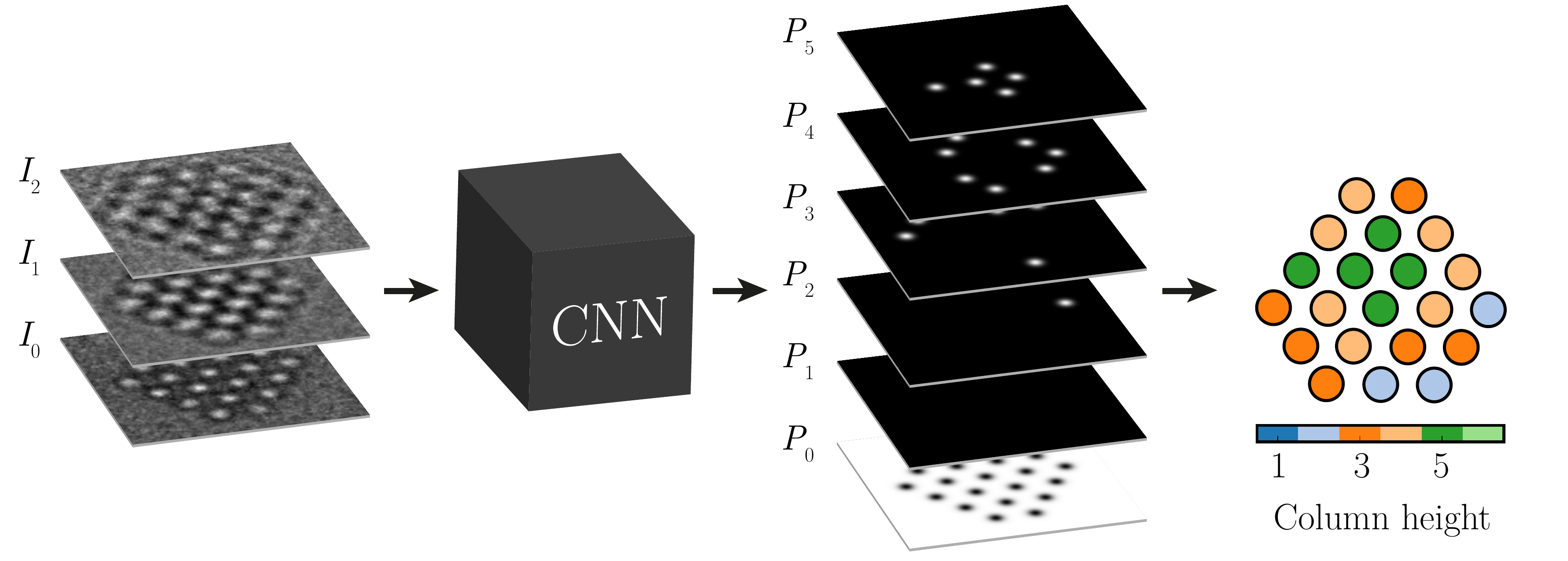}
  \caption{The classification method illustrated on a focal series of
    three images of a gold nanoparticle.  The convolutional
    neural net is fed one or more TEM images of the same sample with
    varying microscope parameters.  The task of the CNN is to classify
    each pixel as belonging to one of six categories: background or an
    atomic column 1--5 atoms high. The output of the CNN is thus six probability
    maps, which are converted into an interpretation of the structure.}
  \label{fig:principle}
\end{figure*}

An example is shown in Fig.~\ref{fig:principle}, where one or more images
of a structure is mapped onto a set of probability maps, from which
the interpreted structure can be depicted.  The input will typically
be a single grey-scale image of size $N_x \times N_y$, but it is
possible to use multiple images of the same spatial region, for
example a focal series where the microscope focus is varied
systematically.  Thus in general the neural network maps image data $I_{x,y,k}$ of shape
$N_x \times N_y \times N_f$ (where $N_f$ is often $1$) to probability
maps $P_{x,y,k}$ of
shape $N_x \times N_y \times N_c$, where $N_c$ is the number of
classes \emph{including a background class}.  Including the background
class makes it easy to enforce normalization of the probabilities,
\begin{equation}
  \label{eq:1}
  \forall x,y: \qquad \sum_k P_{x,y,k} = 1.
\end{equation}

With such a classification scheme, it is important that structures do
not overlap, and overlapping structures should be handled by defining
new classes.  An example is columns of atoms, which can be handled by
making classes for a single atom, a column of two atoms, etc.

\subsection{Preprocessing}
\label{sec:preprocessing}

Contrast and illumination may vary significantly across experimentally
obtained TEM images, in particular if images contain local structures
that are not relevant for the problem being analyzed.  This is handled
by a combination of subtractive and divisive normalization.  First, a local average of the
intensity is subtracted from the image
\begin{equation}
  \label{eq:sub_norm}
  G_{ijk} = I_{ijk} - \frac{1}{N_f} \sum_{pqk'} w_{pq} I_{i+p,j+q,k'}
\end{equation}
where $w_{pq}$ is a Gaussian weighting window normalized so
$\sum_{pq}w_{pq} = 1$.  The decay length of the Gaussian weighting
window must be chosen to be significantly longer than the length
scales of the features the net should detect, to avoid washing them
out.

Finally, the contrast is normalized with a divisive normalization
using the same Gaussian weighting window
\begin{equation}
  \label{eq:div_norm}
  H_{ijk} = \frac{G_{ijk}}{\frac{1}{N_f} \sqrt{\sum_{pqk'} w_{pq} G^2_{i+p,j+q,k'}}}
\end{equation}

\subsection{Neural net architecture}
\label{sec:neur-net-arch}

The neural network needs to be able to combine information on multiple
length scales.  Locally, the atoms are identified as local peaks or
valleys, but estimating what an atom should look like requires
contextual information since it depends on e.g. the defocus of the
microscope.   In some images the atoms may be bright
spots, in other they are dark spots, and the contrast may even invert
within different regions of the same image.

\begin{figure*}[t]
  \centering
  \includegraphics[width=16cm]{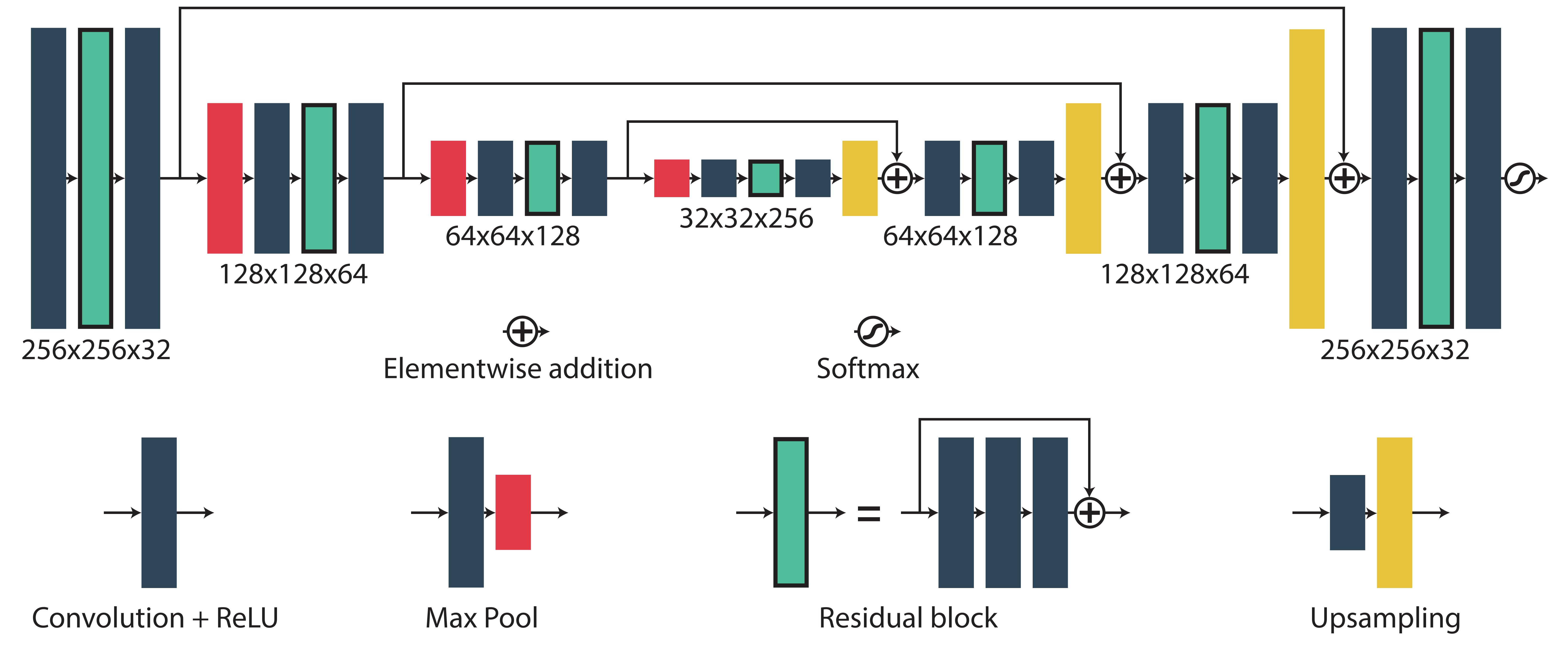}
  \caption{The architecture of the neural network. Information flows
    from left to right. The features are down-sampled in an encoding
    path and up-sampled through a decoding path, in addition several
    skip connections ensure that it is possible to retain fine spatial
    information.}
  \label{fig:architecture}
\end{figure*}

The network architecture is based on fully convolutional networks
(FCN) for pixel-wise segmentation\cite{Shelhamer:2017bc}.  Following the
FusionNet structure proposed by Quan \emph{et al.}
\cite{Quan2016} we  use additive
skips and residual blocks to prevent vanishing gradients and to allow
for training of deeper neural nets.  Multi-level up-sampling and skip
connections combine global abstract information from deep coarse paths
with local spatially resolved information from shallow paths.

The network has a single pipeline with additive skip connections to
preserve spatial information at each resolution.  The lowest
resolution is one eighths of the full resolution, this allows for
spatial filters to be applied that compare features across the entire
image.  The shape of the network is chosen to be symmetric, so for
every layer present in the part where resolution is reduced, there is a
corresponding layer in the part where resolution is increased again.  
The chosen architecture is shown in Fig.~\ref{fig:architecture}

At each resolution on the down-sampling and up-sampling paths, the
network consists of five convolutional layers, with a skip connection
bypassing the middle three layers using elementwise addition (shown as a
\emph{residual block} in Fig.~\ref{fig:architecture}).  Every
convolutional layer except the last employ a $3 \times 3$ convolutional
kernel, followed by an element-wise rectified linear activation,
$h(x) = \max \{0, x\}$, which are then batch normalized following eq.~(\ref{eq:sub_norm}--\ref{eq:div_norm})
\cite{Ioffe2015}.

Feature compression is done in the down-sampling path using a max
pooling layer down-sampling by a factor two in both spatial direction,
while doubling the number of feature maps. Conversely, in the
up-sampling path the features are up-sampled using a transpose
convolutional block\cite{Shelhamer:2017bc} doubling the spatial resolution
while halving the number of feature maps, followed by an element-wise
addition from same level of the encoding path, forming a long skip
connection.

The final scoring consists of a convolutional layer with a $1\times 1$
kernel followed by a softmax non-linearity
\begin{equation}
  \label{eq:4}
  \sigma(P_k) = \frac{\exp(P_k)}{\sum_{k=1}^{m_f} \exp(P_k)} \, .
\end{equation}

The transpose convolutional layers are initialized as bilinear
interpolation and all other layers use random weight initialization.

The network is implemented with TensorFlow using the Python
API,\cite{TensorFlow2016} chosen due to the
wide range of functions already made available, as well as the
community support. All models are trained and tested with TensorFlow
on a single NVIDIA GTX 1080 Ti. Our models and code are publicly
available.\cite{github_jacobjma}

\subsection{Generation of training data}
\label{sec:gener-train-data}

A particular challenge is to generate the training data for the neural
net, since on one hand these data should include the kind of
structures the net should be able to recognize, but on the other hand
should not bias the network towards a specific interpretation of the
images.  This makes it particularly difficult to use real experimental
data as training data, since the network would be trained to reproduce
any subconscious bias of the scientists generating the interpretations
to which the net is trained.

Instead, we train the network to a large set of simulated data.  It is important to be aware that this does not preclude biasing the training set, since such a bias will always be present in the selection criteria generating the structures that form the basis for the image simulations, but at least the true positions of all atoms are known for the simulated images.

We try to minimize the bias of the models by generating a training set with a rather large random component while still maintaining realistic atomic positions, but without resorting to e.g.\ thermodynamical modelling of the systems; this will be discussed further in the sections describing applications of the method.

The training set consists of a collection of computer generated
systems (e.g. nanoparticles, if the neural net is to be applied to
such), generated using the Atomic Simulation Environment (ASE)
\cite{HjorthLarsen:2017hn}.  Simulated images are generated using the
Multislice Algorithm,\cite{Goodman:1974ku}.  Simulation is done
using the publicly available QSTEM code\cite{Koch2002}, through a Python
interface to ASE developed by the authors \cite{Madsen:2017ip}.  The
exit wave functions for each system in the training set is
precomputed, but during training simple symmetry operations
(translation, rotation by 90$^\circ$ and mirroring) can easily be
applied in each training step.

\begin{table*}[tb]
    \centering
    \begin{tabular}{|r|c|c|c|}
    \hline
    parameters & lower bound & upper bound & distribution \\
    \hline
    \hline
    defocus ($\Delta f$) & -200 \AA & 200 \AA & uniform \\
    \hline
    3rd order spherical ($C_s$) & -20 $\mu$m & 20 $\mu$m & uniform \\
    \hline
    5th order spherical ($C_5$) & 0 & 5 mm & uniform \\
    \hline
    1st order astigmatism magnitude & 0 & 100 \AA & uniform \\
    \hline
    1st order astigmatism angle & 0 & 2$\pi$ & uniform \\
    \hline
    deflection \footnote{Blurring caused by noise leading to a random deflection of the image relative to the detector, such as vibrations, drift of the stage, and time-dependent magnetic noise fields resulting from eddy currents in the material of the lenses \cite{Lee:2014bn} } & 0 & 25 \AA & uniform \\
    \hline
    focal spread & 20 \AA & 40 \AA & uniform \\
    \hline
    dose & $10^1$ $e^-/\text{\AA}^2$ & $10^4$ $e^-/\text{\AA}^2$ & exponential \\
    \hline
    $c_1$ (MTF) & 0 & 0.1 & uniform \\
    \hline
    $c_2$ (MTF) & 0.4 & 0.6 & uniform \\
    \hline
    $c_3$ (MTF) & 2 & 3 & uniform \\
    \hline
    \end{tabular}
    \caption{Randomized parameters for generating training examples of graphene for a 80 kV microscope.}
    \label{table:parameters}
\end{table*}

For each training iteration, a Contrast Transfer Function (CTF) is
generated with randomly chosen parameters for the electron microscope
taken from a distribution; Table \ref{table:parameters} shows an
example of parameters used for graphene.  The CTF is then applied to
the precomputed exit wave function.  The effect of energy spread
(i.e.\ temporal coherence) is included in the quasi-coherent
approximation\cite{Kirkland2010}, and temperature effects are included
by blurring the atomic potentials.  The images are resampled to a
random sampling rate, a technique sometimes referred to as
\emph{scale-jittering}.  It is essential to include a reasonable model
of noise in the images, this is done by modelling the finite electron
dose with a Poisson distribution, and including the modulation
transfer function (MTF) of the detector in the image simulation.  The
latter is essential as it has a strong influence on the spectral
properties of the noise, and prevents that the network is
trained incorrectly to detect atoms by the absence of pure white noise.  The MTF
is modelled as \cite{Lee:2014bn}:
\begin{align} \label{eq:mtf}
    \mathrm{MTF}(q) = \frac{1-c_1}{1+\left(\frac{q}{2 c_2 q_N}\right)^{c_3}} + c_1 \quad ,
\end{align}
where $q_N$ is the Nyquist frequency.

The ground truth for the training data is generated as a superposition
of Gaussians with an amplitude of one, centered at the positions of
the atoms (or the mean of the positions for atomic columns).  The
background class is then assigning the remaining probability, such that
the sum of probabilities is one; this is possible since the overlap
between any pair of Gaussians is negligible.  The width, $\sigma$, of
the Gaussians is an important parameter, since it strongly influences
the penalty of wrongly assigning a region of the inferred confidence
map to the background.  We found that a too small value of $\sigma$
would lead to a network with a strong tendency to assign any region
that is difficult to analyze (e.g. due to noise) to the background
class. In a similar way, we found that a common local minimum at
training would be to assign anything to the background class, this is
also exacerbated by a low value of $\sigma$.  We found that a width of
$\sigma = 1$\AA, corresponding to 8--10 pixels at typical resolutions,
worked well for the cases we have considered.

\subsection{Training}

The CNN is trained using a mean squared difference loss
function,\footnote{A good alternative is the cross-entropy loss
  function.  In this case, we find that the difference between the two
is negligible.} regularized with a penalty on the size of the $l_2$ norm of the weights
\begin{align} \label{eq:loss}
    L =  \sum_{ijk} \|\tilde{P}_{ijk} - P_{ijk}\|^2 + \frac{1}{2} \lambda \sum_{i} W_i^2
\end{align}
where $\tilde{P}$ is the output and $P$ is the ground truth.  The network is mainly regularized through the large variability of the training image data, since a new training image is simulated for every training iteration.  Nevertheless, we found that performance on actual experimental data is improved by adding moderate $l_2$ regularization (also known as weight decay), since this causes any weight not being used by the network to produce meaningful output to become negligible rather than to persist for no reason.  Such weights may deteriorate performance on actual experimental data although they do not negatively impact the performance on the training data.

\subsection{Post-processing and interpretation}
\label{sec:interpretation}

While the interpretation of the confidence maps is simplest if there is only a
single class, we here illustrate how it can be done even in the case
of multiple classes.

The first step is identifying the regions where a signal is present.  This
is done by finding all minima in the confidence map for the background
class.  Only minima below $\varepsilon = 0.995$ on a scale
from 0 -- 1 are included, this prevents spending time on analyzing
regions that are clearly background.  The local minima are then used
as seeds for basins created using the watershed principle for image
segmentation using Meyer's algorithm \cite{Dougherty1993}.  We avoid
including long tails in segments by setting a hard upper limit for
each segment at $\varepsilon$.

\begin{figure*}[tp]
  \centering
  \includegraphics[width=16cm]{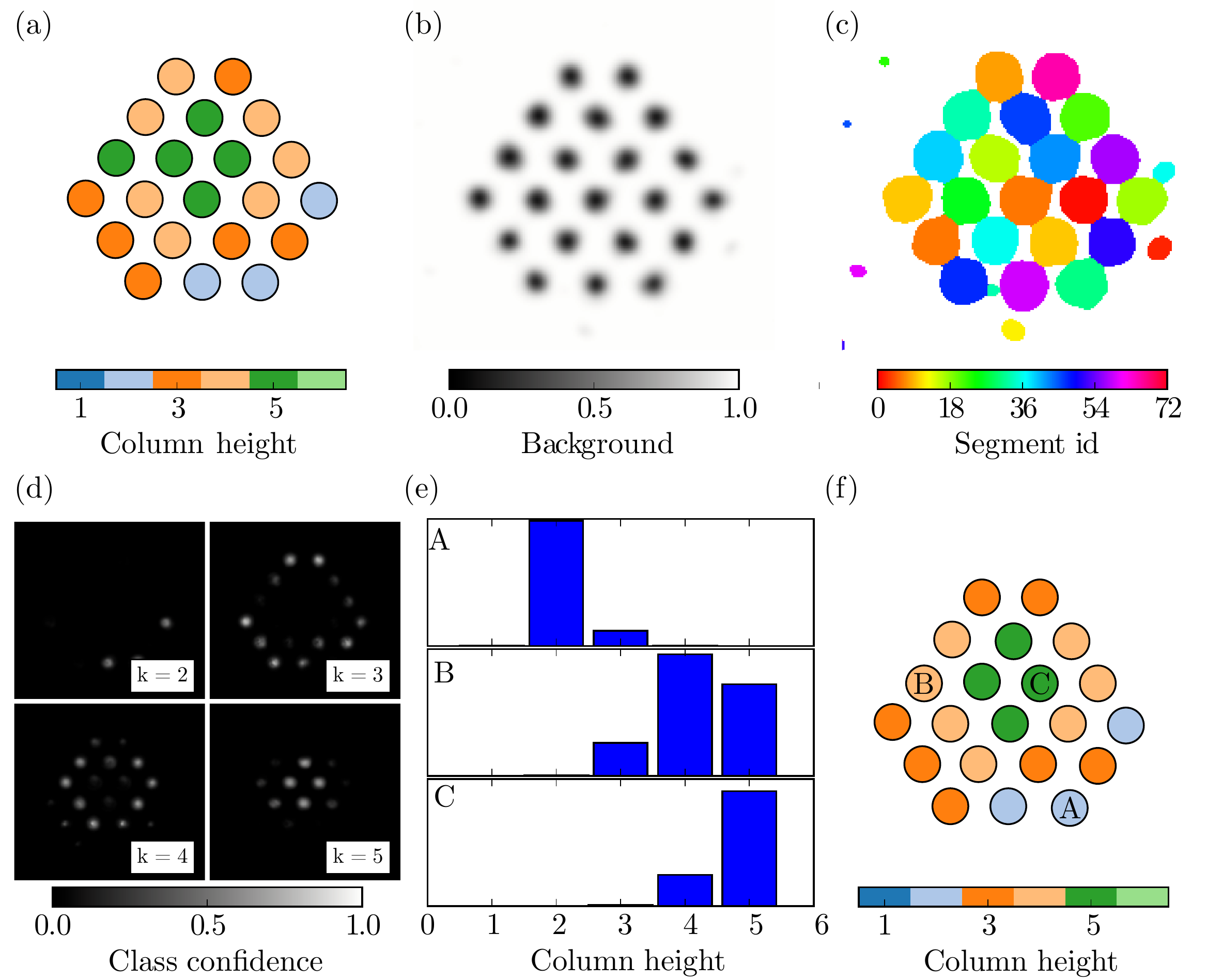}
  \caption{\textbf{(a)}  The input to the network is three images of a
    small nanoparticle, recorded (or simulated) with
    different defocus; here is shown a single simulated image with the
    ground truth thickness of the individual columns marked with
    colors.  \textbf{(b)} The background
    confidence map; we see that the network correctly identifies that
    there is something at each atomic column, but also thinks there
    may be something outside the nanoparticle.  \textbf{(c)} A
    segmentation of the background map into several distinct objects.
    \textbf{(d)}
    Confidence maps for the classes corresponding to columns
    containing two to five atoms.  \textbf{(e)} Each object is
    assigned a probability of belonging to each of the five classes.
    \textbf{(f)} The final classification of the atomic columns.  The
    labels A--C mark the three columns examined in panel (e).  In this
    case, most atomic columns are correctly assigned to their
    classes. Column B which has five atoms is
    incorrectly identified as having only four atoms; however the
    network is clearly in doubt as seen in the probability
    distribution.  One other column is misassigned, in both cases the
    network has probably learned that columns at edges and corners are
    likely to contain fewer atoms, which is not the
    case for these two columns.
  }
  \label{fig:interpretation}
\end{figure*}

Each segment is then assigned a probability for belonging to each non-background class as
\begin{equation}
    p_n(c_k) \propto \sum_{i,j \in \mathcal{S}_n} P_{ijk} \quad k > 0 \quad ,
\end{equation}
where the sum is over all pixels belonging to the $n$'th image
segment. The coordinate of the atomic structure is calculated as the
center-of-mass of the image segment.  Finally, segments are discarded
if
\begin{equation}
\sum_{k>0} p_n(c_k) < t p_n(c_0)\label{eq:3}
\end{equation}
The value chosen for $t$ is normally uncritical, but values near 0.5
are recommended.  It should be noted that in most cases there is only a
single class ($c_1$), ``an atom''.
The process is illustrated in Fig.~\ref{fig:interpretation}.

\section{Application to graphene}

High resolution TEM has been used extensively to study graphene, and several automatic algorithms for extracting quantitative information have been proposed \cite{Kramberger:2016is,Mittelberger:2017in,Vestergaard:2014be}.  Of particular interest is the ability to identify defects, both localized (vacancies, dislocations etc) and extended (grain boundaries).

\subsection{Training}

It is an easy task for a CNN to recognize the regular hexagonal lattice of graphene.  However, we want the network to be able to correctly localize the atomic positions also in situations where they are not at or near their ideal positions.  Thus, the atomic models used to generate the training images cannot simply be ideal sheets of graphene, nor can they be sheets of graphene with added defects.

The opposite extreme, that of generating purely random atomic positions, would result in inefficient networks as the vast majority of the training data would be very different from the experimentally interesting situations.  Instead, we generate atomic positions that lie somewhere between these two extremes.

\begin{figure*}
    \centering
	\includegraphics[clip,width=16cm]{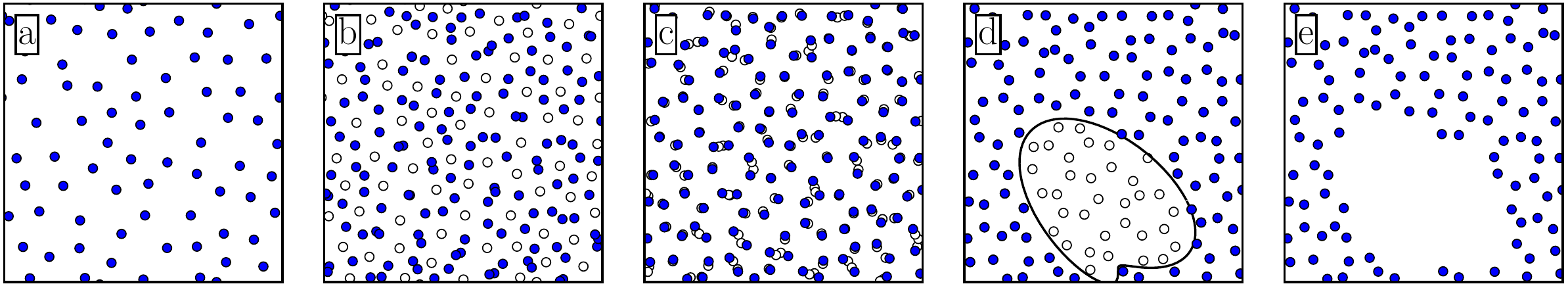}
    \caption{\label{fig:graphene_training} Procedure for generating training structures for graphene. \textbf{(a)} A square in 2D space is filled with randomly distributed seed points under the constraint of a minimum separation in terms of euclidean distance. \textbf{(b)} Next, the Voronoi tessellation is generated from the seed points, the vertices of the diagram will become atomic positions while the original positions are discarded. \textbf{(c)} To avoid overlapping atoms, the positions of the Voronoi vertices are relaxed using Lloyd's algorithm. \textbf{(d)} Lastly, zero to four holes of varying size and shape are introduced in the structure, resulting in the structure seen in \textbf{(e)}.}
\end{figure*}

The algorithm is based on the observation that a Voronoi tessellation of a 2D set of points mainly consists of hexagons, and is illustrated in Fig.~\ref{fig:graphene_training}.  First, an area is filled with randomly distributed points under the constraint of a minimal distance between the points, i.e.\ a Poisson disc distribution.  These points form the generating centers of a Voronoi tessellation; the vertices of the tessellation will become the carbon atoms.  The tessellation is optimized by a few steps of Lloyd's algorithm\cite{LLOYD:1982do}: the centers of the Voronoi tessellation are moved to the center of mass of their respective Voronoi cell.  This makes the Voronoi polyhedra more regular, and in particular it moves closely placed vertices apart, preventing atoms from being placed unrealistically close.  Finally, from zero to four holes are cut randomly in the structure.

\begin{figure*}
    \centering
	\includegraphics[width=16cm]{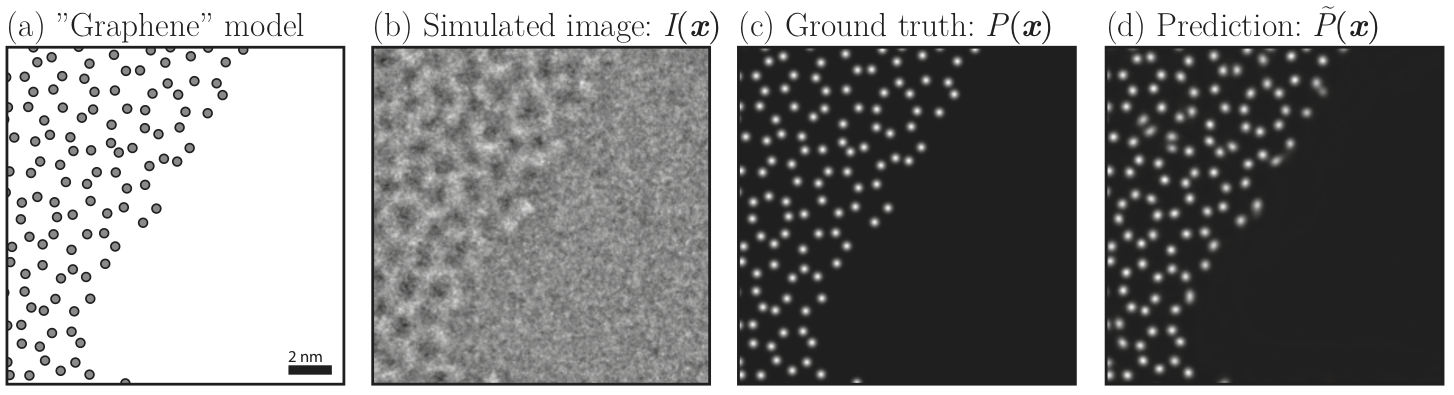}
    \caption{The CNN is trained on simulated images of graphene-like structures, generated by the algorithm in Fig.~\ref{fig:graphene_training}.  \textbf{(a)} A quasi-random graphene-like structure.  \textbf{(b)} Simulated image based on the atomic positions.  \textbf{(c)} The corresponding ground truth calculated from the atomic positions.  The network is trained on a series of matching images and ground truth maps.  \textbf{(d)} The output prediction of the trained neural network given the simulated image in (b).}
    \label{fig:g_training_img} 
\end{figure*}
The resulting structures form a structure which is very suitable for our purpose.  The distribution of bond lengths is quite narrow, and the mean can be controlled by choosing the initial number of points.  The structure contains a large number of polygons with five to eight sides, similar to what is observed in graphene grain boundaries.  A typical training structure is shown in Fig.~\ref{fig:g_training_img}.

We generated 500 random structures with a size of $43.2 \times 43.2$
\AA, or $360 \times 360$ pixels at a sampling rate of $0.12$
\AA/pixel. All the simulations were done at an acceleration voltage of
$80$ kV. While the microscope parameters are uniquely generated at
each training step, the same structure is utilized multiple
times. This have little consequence since most of the variability is
in microscope parameters.

\subsection{Analyzing experimental images}

\begin{figure*}
    \centering
	\includegraphics[width=16cm]{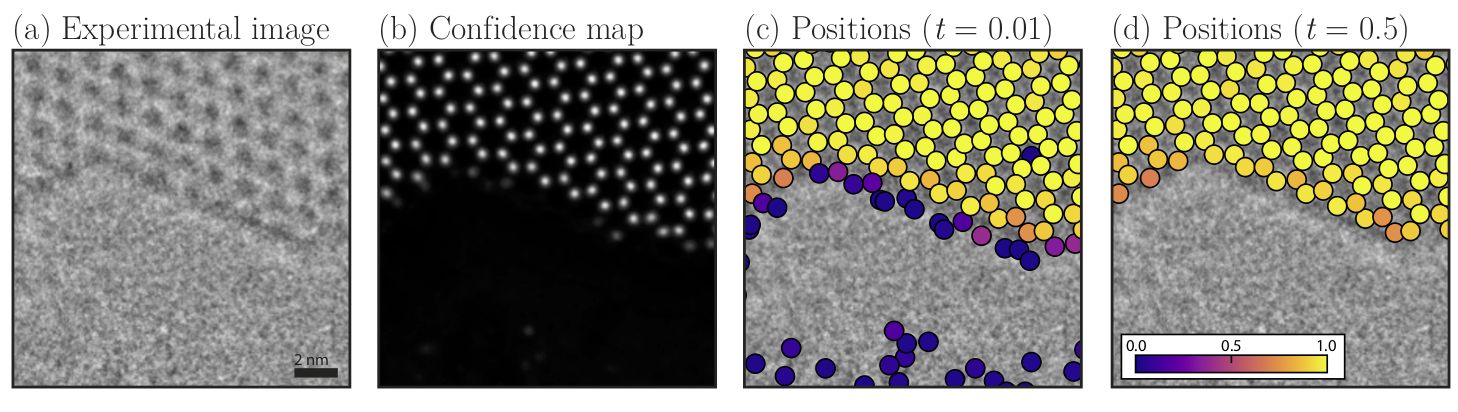}
        \caption{\label{fig:testing} A neural network trained
          exclusively on simulated data is capable of generalizing to
          experimental images. \textbf{(a)} Single suspended graphene
          sheet with a hole formed under the influence of the electron
          beam. \textbf{(b)} The regressed probability distribution
          predicted by the neural network for the image in
          (a). \textbf{(c)} The local peak positions of the
          probability map is overlayed on the image. The peaks are
          color-coded according to their maximum value. Peaks with a
          maximum value less than 0.01 are excluded. (\textbf{d}) A
          higher tolerance for exclusion is used to remove peaks with
          a maximum value less than 0.5.}
\end{figure*}
 An example of how the network performs on experimental data is given
in Fig.~\ref{fig:testing}.  When given an experimental TEM image of
the edge of a graphene sheet, the neural network has no problems
identifying the atoms inside the graphene sheet.  At the edge of the
sheet, there are positions where the network assigns a small but
nonzero probability for the presence of atoms, but using a reasonable
cutoff of $t = 0.5$ gives a result in agreement with a manual analysis
of the image, and without any high-energy atomic configurations at the
edges. 

We apply the trained neural net on a number of graphene images.\cite{Vestergaard:2014be,Kling:2014cx} The experimental graphene
images were measured using a FEI Titan 80-300 Environmental TEM
equipped with a monochromator at the electron gun and spherical
aberration ($C_s$) corrector on the objective lens. The acceleration
voltage of the microscope was 80 kV which is below the knock-on
threshold for carbon atoms in pristine graphene.\cite{Zobelli:2007er}
The electron beam energy spread was below 0.3 eV, while the
$C_s$-corrector was aligned to minimize the spherical aberration. The
images were recorded using a Gatan US1000 CCD camera with an exposure
time of 1 s.

\begin{figure*}[t]
    \centering
	\includegraphics[clip,width=16cm]{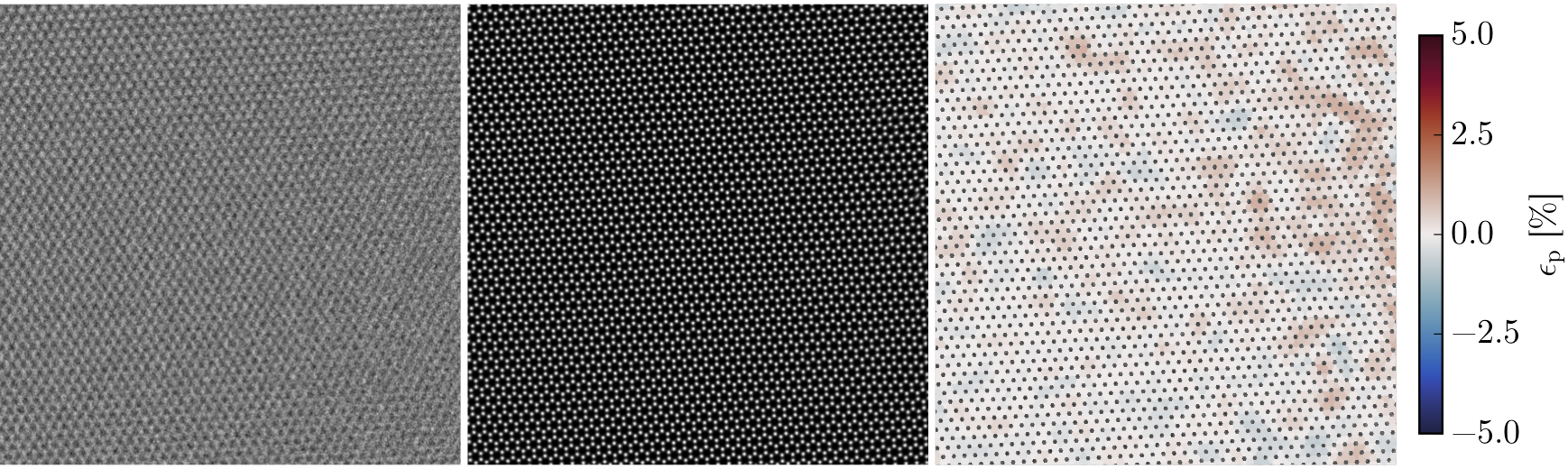}\\
	\includegraphics[clip,width=16cm]{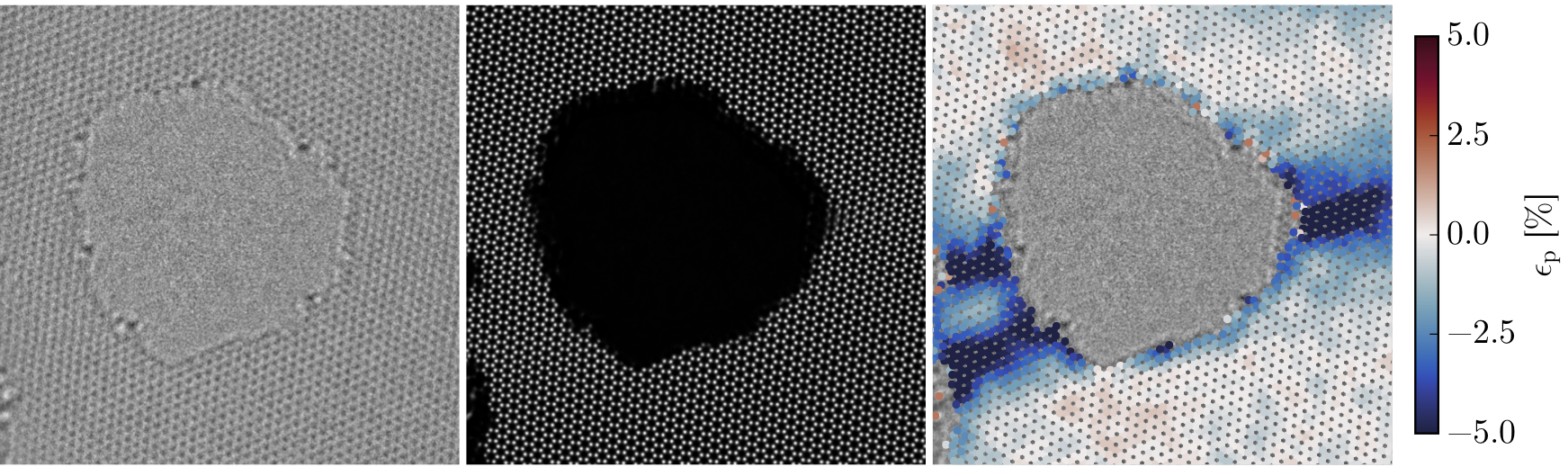}
        \caption{Experimental images of graphene, and their
          interpretation by the neural net.  The first row shows a
          pristine sheet of graphene, the second a sheet with a hole.
          The left column shows the original TEM images.  The center
          column shows the output of the neural net.  The rightmost
          column shows the planar strain calculated from the atomic
          positions, as identified by the neural net.}
    \label{fig:graphene_experimental}
\end{figure*}

Fig. \ref{fig:graphene_experimental} shows a TEM images of pristine
graphene, and of graphene with a hole. The negative $C_s$ imaging
results in images where the carbon atoms are bright spots, with the
centers of the hexagons appearing dark. The output of the neural
network is shown in the central column. The neural network detects all
atomic positions in the pristine sheet, this is accomplished without
having regular hexagonal lattices in the training set. Additionally,
the neural network automatically recognizes that the atoms appear
bright, which is only the case for half of the training images.
Finally, we show the strain calculated from the atomic positions,
using a structural template with the two nearest neighbour shells
(i.e. the 9 nearest neighbours), as described
previously.\cite{Madsen:2017ip}

We do not expect any strain in a pristine area of graphene, hence this sample provides a good of the accuracy of the detected positions. We find that the bond lengths are normally distributed around the mean, 1.42 $\mathrm{\AA}$, with a standard deviation of 0.18 $\mathrm{\AA}$. This standard deviation is comparable or smaller to what was found using the method of Vestergaard et al. \cite{Kling:2014cx}, thus demonstrating, that the positions, determined by our method, are resistant to noise.

\section{Application to metallic nanoparticles}

Metallic nanoparticles on oxide support is a very active research
topic, mainly due to the applications within heterogeneous catalysis.
Often, the detailed atomic structure is important for the catalytic
process, as the active size depending on the process may be e.g.\ step
sites\cite{Honkala:2005iu}, corner atoms\cite{Brodersen:2011jy} or
strained facets\cite{Stephens:2012cx,EscuderoEscribano:2016kf}.

For example, although gold is normally chemically inert, nanoparticles
of gold have been shown to catalyse the oxidation of
CO to CO$_2$\cite{Haruta:1987Novel}.  It is also a system where significant
atomic rearrangement is observed in the presence of gas, both
involving overall shape changes of the
nanoparticles\cite{Uchiyama:2011kq} and changes in the local surface
structure 
We here use supported gold nanoparticles to illustrate
the application of neural nets to the analysis of supported
nanoparticles.

\subsection{Training}

We have trained the network on simulated gold nanoparticles.  As the network should be able to recognize both atomically flat and rough surfaces, the training set includes both kinds of nanoparticles.  Initially, nanoparticles are cut from a regular crystal, keeping a random number of layers in directions with low Miller indices (the $\left<100\right>$, $\left<110\right>$ and $\left<111\right>$ directions).  To roughen the particles, a random number of additional atoms are added to the particle.  The atoms are added at allowed crystal positions at the surface of the nanoparticles, in such a way that highly coordinated surface sites are most likely to be picked.  If the coordination number (i.e. the number of occupied neighbor sites) of site $i$ is $n_i$, then the probability of placing the next atom at site $i$ is chosen as
\begin{equation}
  \label{eq:5}
  p(n_i) = \frac{\exp\left(n_i/T\right)}{\sum_j \exp\left(n_j/T\right)}
\end{equation}
where the sum is over all sites $j$ where $n_j \ge 1$ and $T$ is a parameter that can be chosen differently for each nanoparticle to generate particles with different roughness.

Each particle is then aligned into the  $\left<110\right>$ or $\left<111\right>$ zone axis, and is rotated a random amount around the zone axis.  It is finally tilted 0--5$^{\circ}$ away from the zone axis.

\begin{figure*}[p]
    \centering
    \includegraphics[clip,width=13cm]{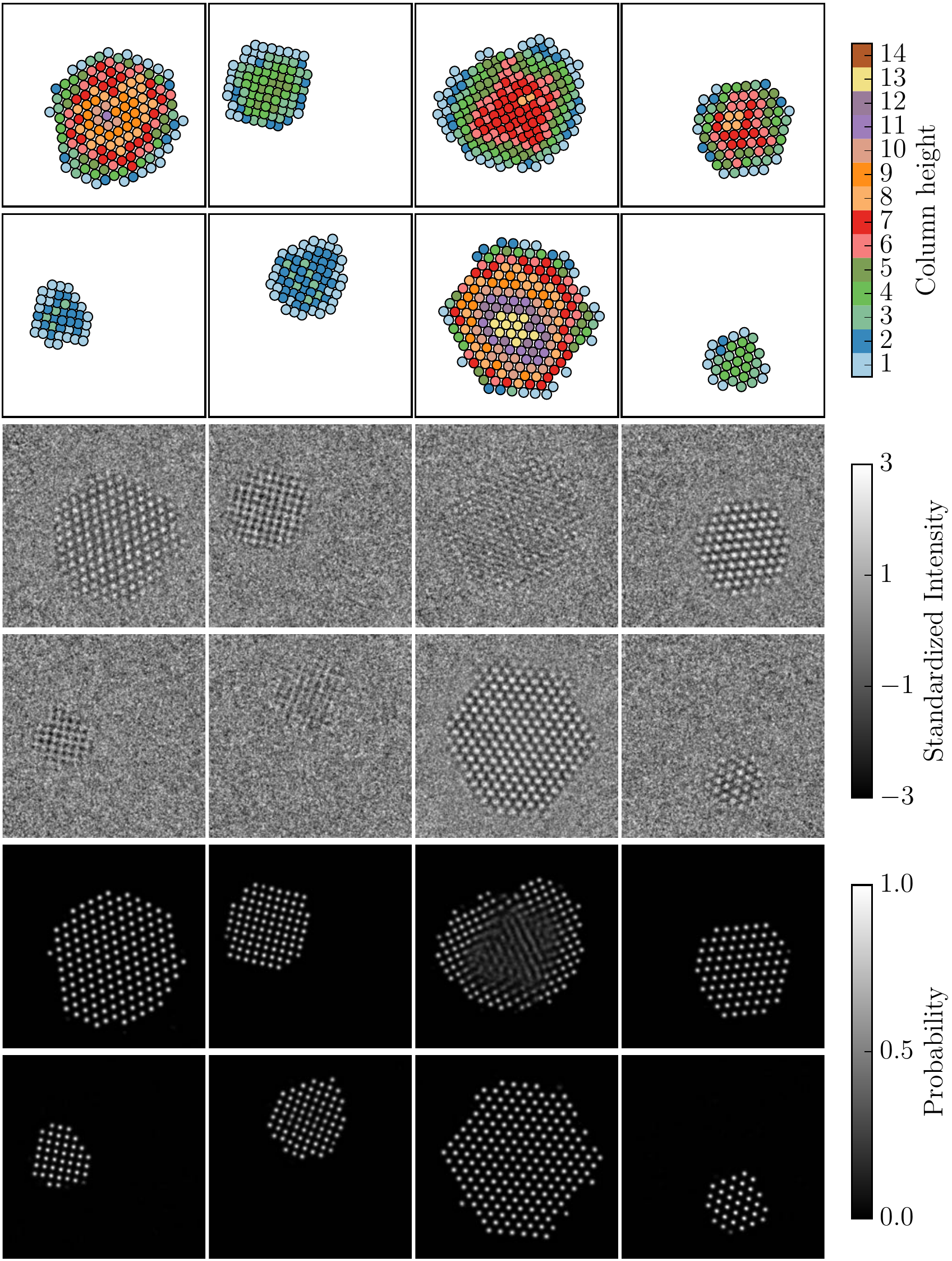}
    \caption{\label{fig:cluster_training} \textbf{Top rows:} Examples of nanoparticles generated using the algorithm we have proposed. The height of the atomic columns of the nanoparticles are indicated with a color-coding. \textbf{Middle rows:} Simulated images given the atomic models above. \textbf{Bottom rows:} Output from our neural network method given the simulated images.  As can be seen, in one of the images the network is not able to identify the atomic columns in the thickest part of the nanoparticle, this is due to a combination of low signal-to-noise ratio in the image, and a significant off-axis tilt smearing out the highest atomic columns.}
  \end{figure*}

As was the case for the graphene simulations, 500  nanoparticles
were generated, but during the training new microscope parameters were
picked for each iteration, and the nanoparticles were randomly
translated, mirrored and rotated by a multiple of 90$^\circ$
(operations that can cheaply be performed on the precomputed wave-functions).  Figure \ref{fig:cluster_training} shows a sample of generated nanoparticles, and their corresponding images.

\subsection{Analyzing nanoparticle image sequence}
\label{sec:analyz-nanop-imag}

\begin{figure*}[t]
  \centering
  \includegraphics[clip,width=13cm]{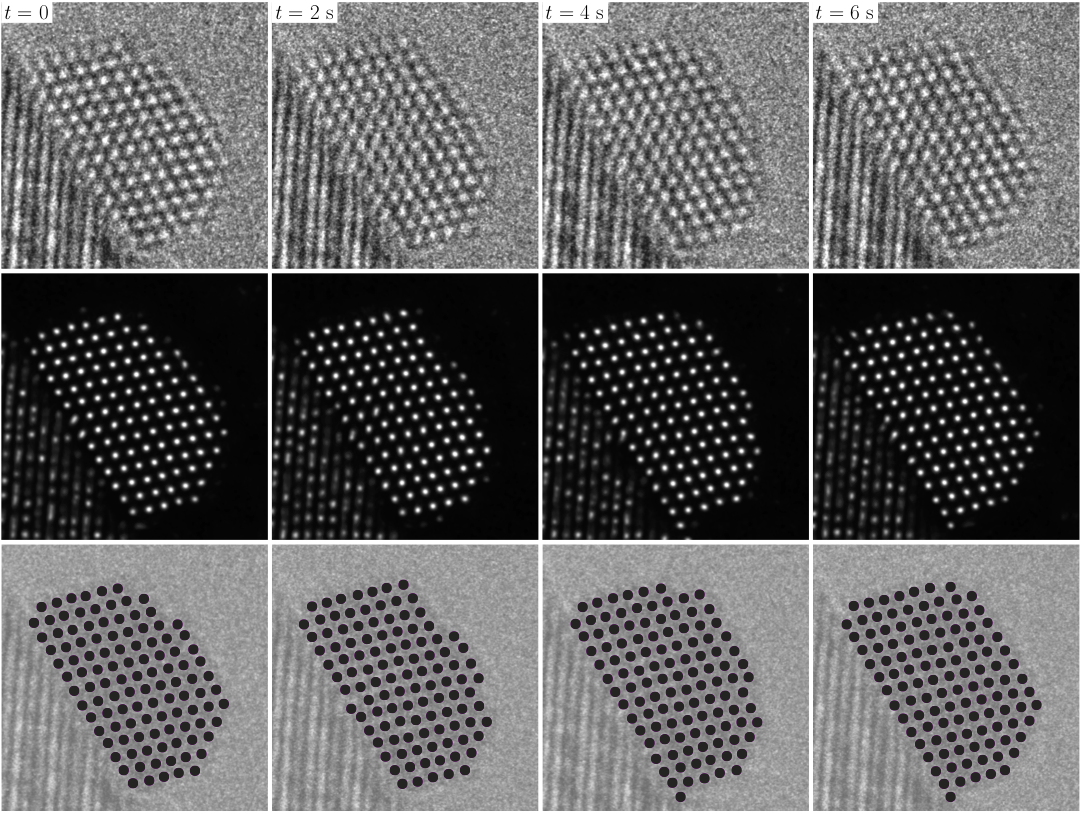}
  \caption{\label{fig:nanoparticle} \textbf{top row:} Experimental
    high resolution
    TEM image of gold on ceria in an oxygen atmosphere. The pressure
    was 4.5 Pa and the electron dose per image was $3.1 \times 10^2$
    e$^-$/\AA$^2$ (dose rate $1.56 \times 10^3$  e$^-$/\AA$^2$/s at
    an exposure time of 0.2 s). 
    \textbf{Middle row:} Output from the neural network given the
    images above. \textbf{Bottom row:} The atoms identified by the
    neural net are marked as purple circles overlaid on the original
    image.  The experimental images were measured using a FEI Titan
    80-300 Environmental TEM operated at 300 keV.}
\end{figure*}

We applied the resulting network to gold nanoparticles on a ceria substrate.  Figure~\ref{fig:nanoparticle} shows a TEM image of such a particle, and the corresponding analysis by the neural net.  It is seen that the network confidently identifies the atoms in the nanoparticle, but not in the substrate; this is partly due to the network not being trained on ceria's crystal structure, partly because the substrate is not in a prominent zone-axis orientation.

In the microscope, a image sequence of this nanoparticle was recorded,
Fig.~\ref{fig:nanoparticle} shows four snapshots of this sequence,
clearly showing the atomic diffusion processes.

\begin{figure*}[tb]
  \centering
  \includegraphics[clip,width=13cm]{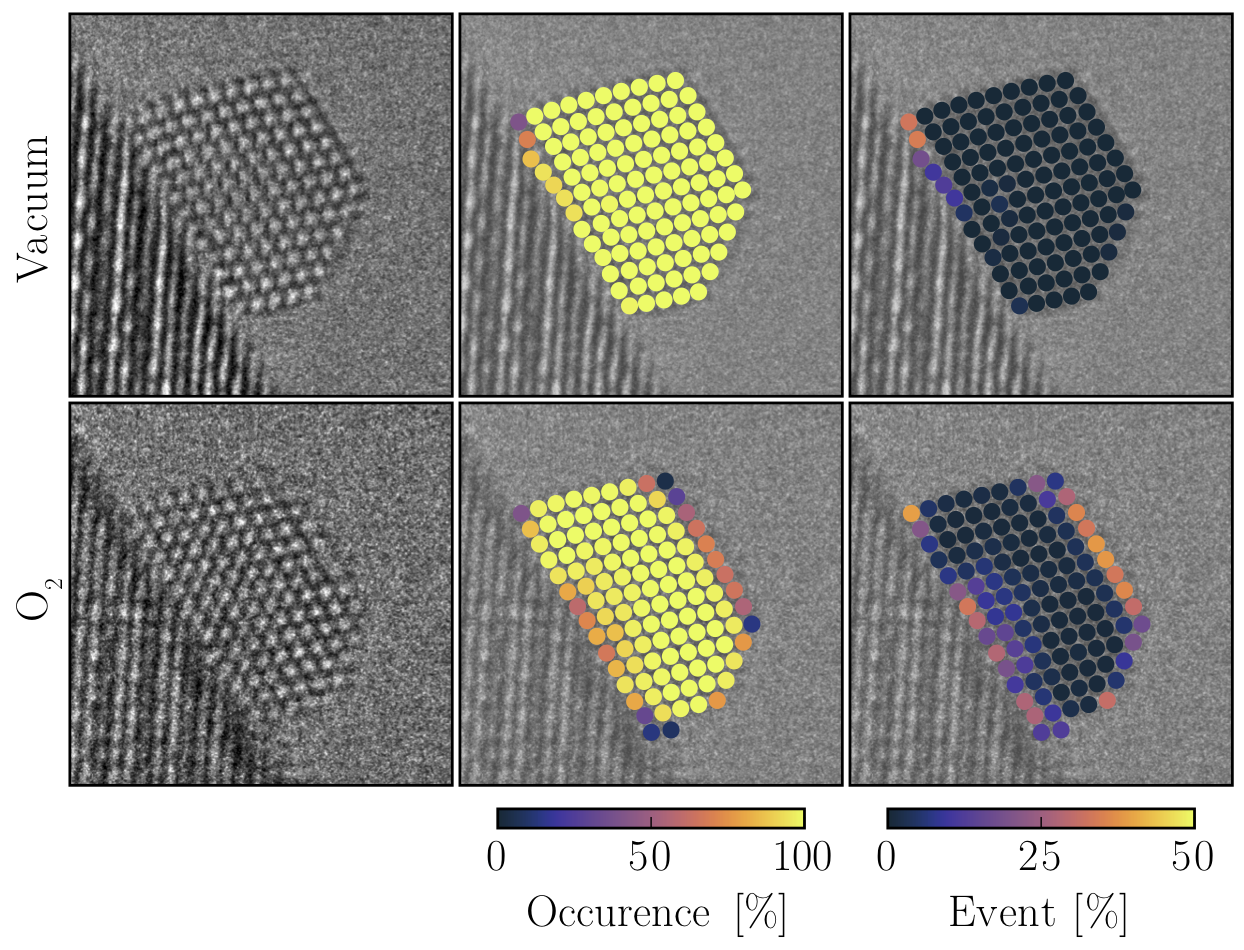}
  \caption{\label{fig:gasses} The surface dynamics of gold
    nanoparticles is influenced by the gaseous atmosphere. The
    occurence is the percentage of frames where the neural network
    identified an atomic column at a possible site. And the events are
    the percentage of frames where a site was previously occupied, but
    is unoccupied in the frame immediately after, or vice versa.}
\end{figure*}
  
We used the neural network to analyze TEM image sequences showing surface
diffusion on gold nanoparticle in various gasses.
Figure~\ref{fig:gasses} shows the same ceria-supported gold
nanoparticle in high vacuum and in an oxygen atmosphere.  The neural
network is applied to each frame in the image sequence, and used to
identify the presence (and position) of the atomic columns.  Since the
network was not trained on substrates, and since atomic resolution may not be
obtainable simultaneously in the substrate and the
nanoparticle, we only use the network to analyze the metallic
nanoparticle and mask out the output of the network corresponding to the substrate.

During the image sequence, atoms at the surfaces and in particular at the corners of the nanoparticle are clearly seen to appear and disappear again, as the surface atoms diffuse on the particle.  We illustrate this in the figure in two different ways.  In the middle column, atomic columns are colored according to the fraction of time they are present in the image.  It is clearly seen that in the presence of oxygen, many of the surface and corner atoms are only present part of the time, indicating surface diffusion.  In the rightmost columns, ``events'' are counted.  It is considered an ``event'' if an atomic column is present in one frame, but absent in the next, or \emph{vice versa}; this cause the diffusing atoms to light up on the figure.  Together, this analysis shows that the presence of oxygen clearly enhances the surface diffusion.  This will be the discussed a separate publication. 

\subsection{Counting atomic columns}

We envision that the main application of this technique will be to
identify atoms or columns of atoms, as demonstrated in the two
afore-mentioned examples.  For this kind of applications, it is
relatively straight-forward to train a neural net on sets of simulated
images, such that the net becomes able to identify the positions of
the atoms or atomic columns also in experimentally obtained images.

A far more demanding task is to identify the chemical species of
single atoms in 2D materials, or to count the number of atoms in
atomic columns in nanoparticles.  It does not
appear to be possible to train a network that solves this kind of
tasks based on a single image.  However, using a
series of images taken with varying focus settings, it appears to be
possible to train such networks to identify multiple mutually
exclusive atomic objects, as already hinted in
Fig.~\ref{fig:interpretation}.

The most common method for measuring the height of atomic columns is
Scanning Transmission Electron Microscopy (STEM)\cite{Jones:2014gk,DeBacker:2017fs}.  In
TEM, the peak intensity is not generally a monotonic function of
column heights, and is sensitive to small tilts of the sample.
However, Gonnissen \emph{et al.}\cite{Gonnissen:2017dh} show that in
principle the peak profile in TEM contains enough information to
reconstruct the height of the atomic column, and that greater accuracy
can be obtained with TEM than with STEM, provided that information
from all pixels are used and not just the peak intensities.  However,
no practical demonstration of such a model exist.  It has, however,
been demonstrated that the number of atoms in a column can be counted
by reconstructing the exit wave function from a focal series \cite{Chen:2015ik}.

To address this problem, we have trained a neural net to classify the
columns by the number of atoms.  The net is trained on focal series of
three simulated TEM images.  While the absolute defocus cannot be
known precisely in practise, it is relatively easy to control the
difference in defocus between the images in a focal series.  For this
reason, we train the network on three images with defocus of $-170
\,\text{\AA}+ \Delta$,  $-40 \,\text{\AA}+ \Delta$ and  $90
\,\text{\AA}+ \Delta$, where $\Delta$ is a random number in the range
$-20\,\text{\AA}$ -- $+20\,\text{\AA}$.  The sperical aberration was
varied between $C_s = -10 \mu m$ and  $C_s = -20 \mu m$, and all other
parameters as reported previously.

\begin{figure*}
  \centering
    \includegraphics[clip,width=13cm]{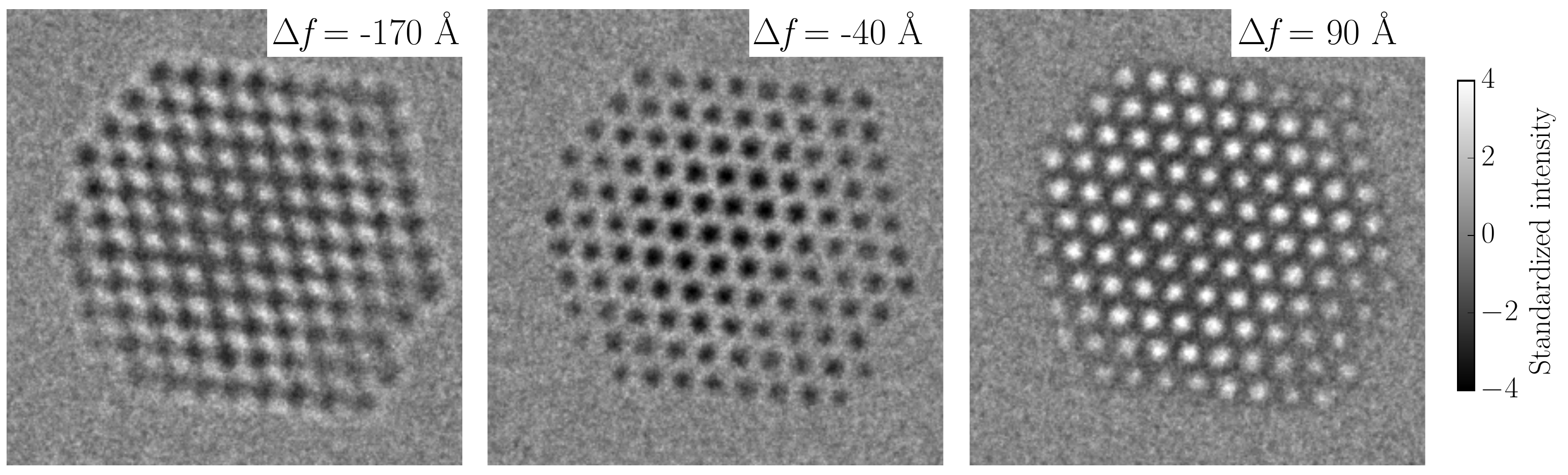}
    \includegraphics[clip,width=13cm]{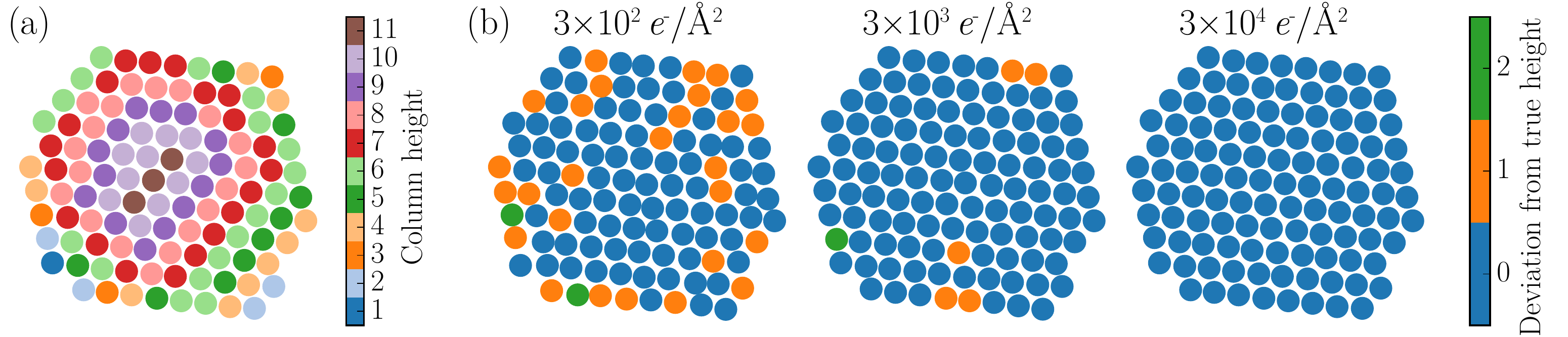}
  \caption{Top row: An example of the input images for the neural net,
    simulated at three values of the defocus.  Bottom row:  The
    performance of the neural net.  (a) shows the ground truth,
    i.e. the actual number of atoms in the atomic columns.  (b) shows
    the error made by the neural network at three different electron
    doses.  The dose given is the sum of the dose in the three
    images.}
  \label{fig:counting}
\end{figure*}

\begin{figure}
  \centering
  \includegraphics[clip,width=8cm]{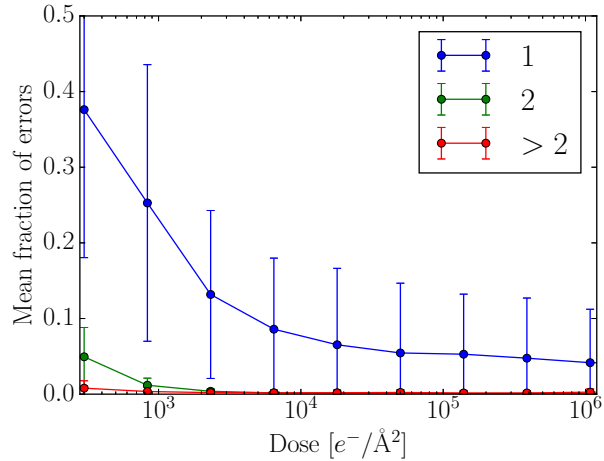}
  \caption{The error rate as a function of electron dose.  The three
    lines show the fraction of atomic columns miscounted by one, two
    or at least three atoms.  The dose is a total for all three
    images.}
  \label{fig:countgraph}
\end{figure}

An example of the performance is given in Fig.~\ref{fig:counting}.  We
see that at the highest dose of $3 \times 10^4 e^-/\text{\AA}^2$,
there are no errors. With a dose a 100 times smaller, a significant
number of errors appear, but almost all of them are only one-off.  It
should, however, be noted that this result is obtained under 
ideal conditions, as the nanoparticle is almost  exactly in the
$\left<110\right>$ zone axis.  When nanoparticles are tilted by up to
$3^{\circ}$, the performance is significantly worse, although most
errors are still only of a single atom.  Figure~\ref{fig:countgraph} shows the fraction
of errors as a function of electron dose for an ensemble of
nanoparticles tilted up to $3^{\circ}$ from the zone axis.

We have not verified the viability of the atom counting method on actual
experimental TEM images.  Doing so would require a significant
experimental effort to obtain reliable atomic column heights either
from STEM or electron tomography on the same nanoparticles used for
the neural net analysis.  Such an effort is beyond the scope of this
publication, but may be the subject of later reports.

\section{Conclusion}
\label{sec:conclusion}

We have demonstrated that deep convolutional neural networks can be
trained to recognize the local atomic structure in High Resolution
Transmission Electron Microscopy images.  The network can be trained
entirely on simulated data, but is capable of giving interpretations
of experimental images that match those of a trained microscopist.  We
have demonstrated the method both on single layers of defected
graphene, and on nanoparticles of gold on a cerium oxide substrate.
In addition, we show that neural nets have the potential to determine
the height of atomic columns from TEM images.

\section*{Acknowledgments}
\label{sec:acknowledgments}

We gratefully acknowledge funding through grant 1335-00027B from the
Danish Council for Independent Research.

\bibliographystyle{local}
\bibliography{references,local}

\begin{thebibliography}{10}
\providecommand{\url}[1]{\texttt{#1}}
\providecommand{\urlprefix}{URL }

\bibitem{McMullan:2014jn}
G.~McMullan, A.~R. Faruqi, D.~Clare, R.~Henderson.
\newblock \emph{Ultramicroscopy}  \textbf{2014}.
\newblock \emph{147}, 156.

\bibitem{Wagner:2012ec}
J.~B. Wagner, F.~Cavalca, C.~D. Damsgaard, L.~D.~L. Duchstein, T.~W. Hansen.
\newblock \emph{Micron}  \textbf{2012}.
\newblock \emph{43}, 1169.

\bibitem{Meyer:2011ez}
J.~C. Meyer, S.~Kurasch, H.~J. Park, V.~Skakalova, D.~K{\"u}nzel, A.~Gross,
  A.~Chuvilin, G.~Algara-Siller, S.~Roth, T.~Iwasaki, U.~Starke, J.~H. Smet,
  U.~Kaiser.
\newblock \emph{Nat. Mater.}  \textbf{2011}.
\newblock \emph{10}, 209.

\bibitem{Meyer:2008hj}
J.~C. Meyer, C.~Kisielowski, R.~Erni, M.~D. Rossell, M.~F. Crommie, A.~Zettl.
\newblock \emph{Nano Lett.}  \textbf{2008}.
\newblock \emph{8}, 3582.

\bibitem{He:2014ew}
X.~He, T.~Xu, X.~Xu, Y.~Zeng, J.~Xu, L.~Sun, C.~Wang, H.~Xing, B.~Wu, A.~Lu,
  D.~Liu, X.~Chen, J.~Chu.
\newblock \emph{Sci. Rep.}  \textbf{2014}.
\newblock \emph{4}, 6544.

\bibitem{Nagao:2015jm}
K.~Nagao, T.~Inuzuka, K.~Nishimoto, K.~Edagawa.
\newblock \emph{Phys. Rev. Lett.}  \textbf{2015}.
\newblock \emph{115}, 075501.

\bibitem{Li:2017ke}
X.~Li, S.~Cheng, S.~Deng, X.~Wei, J.~Zhu, Q.~Chen.
\newblock \emph{Sci. Rep.}  \textbf{2017}.
\newblock \emph{7}, 40911.

\bibitem{Schneider:2014do}
S.~Schneider, A.~Surrey, D.~Pohl, L.~Schultz, B.~Rellinghaus.
\newblock \emph{Micron}  \textbf{2014}.
\newblock \emph{63}, 52.

\bibitem{Hussaini:2017hr}
Z.~Hussaini, P.~A. Lin, B.~Natarajan, W.~Zhu, R.~Sharma.
\newblock \emph{Ultramicroscopy}  \textbf{2017}.
\newblock \emph{186}, 139.

\bibitem{Hytch:1998gr}
M.~J. H{\"y}tch, E.~Snoeck, R.~Kilaas.
\newblock \emph{Ultramicroscopy}  \textbf{1998}.
\newblock \emph{74}, 131.

\bibitem{Zhu:2013er}
Y.~Zhu, C.~Ophus, J.~Ciston, H.~Wang.
\newblock \emph{Acta Mater.}  \textbf{2013}.
\newblock \emph{61}, 5646.

\bibitem{Madsen:2017ip}
J.~Madsen, P.~Liu, J.~B. Wagner, T.~W. Hansen, J.~Schi{\o}tz.
\newblock \emph{Adv. Struct. Chem. Imag.}  \textbf{2017}.
\newblock pp. 1--12.

\bibitem{Bierwolf:1993ic}
R.~Bierwolf, M.~Hohenstein, F.~Phillipp, O.~Brandt, G.~E. Crook, K.~Ploog.
\newblock \emph{Ultramicroscopy}  \textbf{1993}.
\newblock \emph{49}, 273.

\bibitem{Galindo:2007jj}
P.~L. Galindo, S.~Kret, A.~M. Sanchez, J.-Y. Laval, A.~Y{\'a}{\~n}ez,
  J.~Pizarro, E.~Guerrero, T.~Ben, S.~I. Molina.
\newblock \emph{Ultramicroscopy}  \textbf{2007}.
\newblock \emph{107}, 1186.

\bibitem{Zuo:2014fw}
J.-M. Zuo, A.~B. Shah, H.~Kim, Y.~Meng, W.~Gao, J.-L. Rouvi{\'e}re.
\newblock \emph{Ultramicroscopy}  \textbf{2014}.
\newblock \emph{136}, 50.

\bibitem{Zhu:2016qoi}
Y.~Zhu, Q.~Ouyang, Y.~Mao.
\newblock \emph{BMC Bioinformatics}  \textbf{2017}.
\newblock \emph{18}, 348.

\bibitem{Quan2016}
T.~M. Quan, D.~G.~C. Hildebrand, W.-K. Jeong.
\newblock \emph{arXiv}  \textbf{2016}.
\newblock p. 1612.05360.

\bibitem{Kirschner:2000kb}
H.~Kirschner, R.~Hillebrand.
\newblock \emph{Inf. Sci.}  \textbf{2000}.
\newblock \emph{129}, 31.

\bibitem{Meyer:2008je}
{Meyer}, {Heindl}.
\newblock \emph{J. Microsc.}  \textbf{2008}.
\newblock \emph{191}, 52.

\bibitem{Ziatdinov:2017ct}
M.~Ziatdinov, O.~Dyck, A.~Maksov, X.~Li, X.~Sang, K.~Xiao, R.~R. Unocic,
  R.~Vasudevan, S.~Jesse, S.~V. Kalinin.
\newblock \emph{ACS nano}  \textbf{2017}.
\newblock \emph{11}, 12742.

\bibitem{Shelhamer:2017bc}
E.~Shelhamer, J.~Long, T.~Darrell.
\newblock \emph{IEEE Trans. Pattern Anal. Mach. Intell.}  \textbf{2017}.
\newblock \emph{39}, 640.

\bibitem{Ioffe2015}
S.~Ioffe, C.~Szegedy.
\newblock \emph{arXiv}  \textbf{2015}.
\newblock p. 1502.03167.

\bibitem{TensorFlow2016}
M.~Abadi, P.~Barham, J.~Chen, Z.~Chen, A.~Davis, J.~Dean, M.~Devin,
  S.~Ghemawat, G.~Irving, M.~Isard, M.~Kudlur, J.~Levenberg, R.~Monga,
  S.~Moore, D.~G. Murray, B.~Steiner, P.~Tucker, V.~Vasudevan, P.~Warden,
  M.~Wicke, Y.~Yu, X.~Zheng.
\newblock In \emph{Proceedings of the 12th USENIX Symposium on Operating
  Systems Design and Implementation (OSDI {\textquoteright}16)} \textbf{2016} .

\bibitem{github_jacobjma}
{https://github.com/jacobjma}.

\bibitem{HjorthLarsen:2017hn}
A.~Hjorth~Larsen, J.~J{\o}rgen~Mortensen, J.~Blomqvist, I.~E. Castelli,
  R.~Christensen, M.~Du{\l}ak, J.~Friis, M.~N. Groves, B.~Hammer, C.~Hargus,
  E.~D. Hermes, P.~C. Jennings, P.~Bjerre~Jensen, J.~Kermode, J.~R. Kitchin,
  E.~Leonhard~Kolsbjerg, J.~Kubal, K.~Kaasbjerg, S.~Lysgaard,
  J.~Bergmann~Maronsson, T.~Maxson, T.~Olsen, L.~Pastewka, A.~Peterson,
  C.~Rostgaard, J.~Schi{\o}tz, O.~Sch{\"u}tt, M.~Strange, K.~S. Thygesen,
  T.~Vegge, L.~Vilhelmsen, M.~Walter, Z.~Zeng, K.~W. Jacobsen.
\newblock \emph{J Phys Condens Matter}  \textbf{2017}.
\newblock \emph{29}, 273002.

\bibitem{Goodman:1974ku}
P.~Goodman, A.~F. Moodie.
\newblock \emph{Acta Crystallogr. A}  \textbf{1974}.
\newblock \emph{30}, 280.

\bibitem{Koch2002}
C.~Koch.
\newblock \emph{{Determination of core structure periodicity and point defect
  density along dislocations}}.
\newblock Ph.D. thesis, Arizona State University \textbf{2002}.

\bibitem{Lee:2014bn}
Z.~Lee, H.~Rose, O.~Lehtinen, J.~Biskupek, U.~Kaiser.
\newblock \emph{Ultramicroscopy}  \textbf{2014}.
\newblock \emph{145}, 3.

\bibitem{Kirkland2010}
E.~J. Kirkland.
\newblock \emph{Advanced Computing in Electron Microscopy}.
\newblock Springer US, 978-1-4419-6532-5, 2 edn. \textbf{2010}.

\bibitem{Dougherty1993}
E.~R. Dougherty.
\newblock \emph{{Mathematical morphology in image processing}}.
\newblock M. Dekker, New York \textbf{1993}.

\bibitem{Kramberger:2016is}
C.~Kramberger, J.~C. Meyer.
\newblock \emph{Ultramicroscopy}  \textbf{2016}.
\newblock \emph{170}, 60.

\bibitem{Mittelberger:2017in}
A.~Mittelberger, C.~Kramberger, C.~Hofer, C.~Mangler, J.~C. Meyer.
\newblock \emph{Microsc. Microanal.}  \textbf{2017}.
\newblock \emph{23}, 809.

\bibitem{Vestergaard:2014be}
J.~S. Vestergaard, J.~Kling, A.~B. Dahl, T.~W. Hansen, J.~B. Wagner, R.~Larsen.
\newblock \emph{Microsc. Microanal.}  \textbf{2014}.
\newblock \emph{20}, 1772.

\bibitem{LLOYD:1982do}
S.~P. Lloyd.
\newblock \emph{IEEE Trans. Inf. Theory}  \textbf{1982}.
\newblock \emph{28}, 129.

\bibitem{Kling:2014cx}
J.~Kling, J.~S. Vestergaard, A.~B. Dahl, N.~Stenger, T.~J. Booth,
  P.~B{\o}ggild, R.~Larsen, J.~B. Wagner, T.~W. Hansen.
\newblock \emph{Carbon}  \textbf{2014}.
\newblock \emph{74}, 363.

\bibitem{Zobelli:2007er}
A.~Zobelli, A.~Gloter, C.~P. Ewels, G.~Seifert, C.~Colliex.
\newblock \emph{Phys. Rev. B}  \textbf{2007}.
\newblock \emph{75}, 245402.

\bibitem{Honkala:2005iu}
K.~Honkala, A.~Hellman, I.~N. Remediakis, A.~Logadottir, A.~Carlsson, S.~Dahl,
  C.~H. Christensen, J.~K. N{\o}rskov.
\newblock \emph{Science}  \textbf{2005}.
\newblock \emph{307}, 555.

\bibitem{Brodersen:2011jy}
S.~H. Brodersen, U.~Gr{\o}nbjerg, B.~Hvolb{\ae}k, J.~Schi{\o}tz.
\newblock \emph{J. Catal.}  \textbf{2011}.
\newblock \emph{284}, 34.

\bibitem{Stephens:2012cx}
I.~E.~L. Stephens, A.~S. Bondarenko, U.~Gr{\o}nbjerg, J.~Rossmeisl,
  I.~Chorkendorff.
\newblock \emph{Energy Environ. Sci.}  \textbf{2012}.
\newblock \emph{5}, 6744.

\bibitem{EscuderoEscribano:2016kf}
M.~Escudero-Escribano, P.~Malacrida, M.~H. Hansen, U.~G. Vej-Hansen,
  A.~Velazquez-Palenzuela, V.~Tripkovic, J.~Schi{\o}tz, J.~Rossmeisl, I.~E.~L.
  Stephens, I.~Chorkendorff.
\newblock \emph{Science}  \textbf{2016}.
\newblock \emph{352}, 73.

\bibitem{Haruta:1987Novel}
M.~Haruta, T.~Kobayashi, H.~Sano, N.~Yamada.
\newblock \emph{Chem. Lett.}  \textbf{1987}.
\newblock \emph{16}, 405.

\bibitem{Uchiyama:2011kq}
T.~Uchiyama, H.~Yoshida, Y.~Kuwauchi, S.~Ichikawa, S.~Shimada, M.~Haruta,
  S.~Takeda.
\newblock \emph{Angew. Chem. Int. Ed. Engl.}  \textbf{2011}.
\newblock \emph{50}, 10157.

\bibitem{Jones:2014gk}
L.~Jones, K.~E. MacArthur, V.~T. Fauske, A.~T.~J. van Helvoort, P.~D. Nellist.
\newblock \emph{Nano Lett.}  \textbf{2014}.
\newblock \emph{14}, 6336.

\bibitem{DeBacker:2017fs}
A.~De~Backer, L.~Jones, I.~Lobato, T.~Altantzis, B.~Goris, P.~D. Nellist,
  S.~Bals, S.~Van~Aert.
\newblock \emph{Nanoscale}  \textbf{2017}.
\newblock \emph{9}, 8791.

\bibitem{Gonnissen:2017dh}
J.~Gonnissen, A.~De~Backer, A.~J. den Dekker, J.~Sijbers, S.~Van~Aert.
\newblock \emph{Ultramicroscopy}  \textbf{2017}.
\newblock \emph{174}, 112.

\bibitem{Chen:2015ik}
F.~R. Chen, C.~Kisielowski, D.~Van~Dyck.
\newblock \emph{Micron}  \textbf{2015}.
\newblock \emph{68}, 59.

\end{thebibliography}

\end{document}